\documentclass[12pt]{article}
\newcommand{\blind}{0}

\usepackage{booktabs, subcaption, multirow} 
\usepackage{authblk}
\usepackage{url}
\usepackage{array}
\usepackage{natbib} 
\usepackage{amsthm,amsfonts}
\usepackage{amsmath,amssymb} 
\usepackage{color}
\usepackage{bm}
\usepackage{graphicx}
\usepackage[most]{tcolorbox} 

\newcommand{\revb}[1]{{\color{black} #1}} 
\definecolor{lightgrayblue}{RGB}{235, 242, 250}  
\definecolor{darkgrayblue}{RGB}{70, 90, 110}     
\definecolor{lightbeige}{RGB}{252, 248, 237}     
\definecolor{darkbeige}{RGB}{120, 100, 80}       

\newtcolorbox[auto counter]{promptBox}[2][]{%
  colback=lightgrayblue,
  colframe=darkgrayblue,
  coltext=black,
  fonttitle=\bfseries,
  title=Prompt~\thetcbcounter: #2,
  sharp corners=south,  
  boxrule=0.75pt,       
  enhanced
}

\newtcolorbox[auto counter]{outputBox}[2][]{%
  colback=lightbeige,
  colframe=darkbeige,
  coltext=black,
  fonttitle=\bfseries,
  title=ChatGPT's Output~\thetcbcounter: #2,
  sharp corners=south,  
  boxrule=0.75pt,       
  enhanced
}

\newtcolorbox[auto counter]{outputBoxGemini}[2][]{%
  colback=lightbeige,
  colframe=darkbeige,
  coltext=black,
  fonttitle=\bfseries,
  title=Gemini's Output~A\thetcbcounter: #2,
  sharp corners=south,  
  boxrule=0.75pt,       
  enhanced
}

\begin{document}
\def\spacingset#1{\renewcommand{\baselinestretch}%
{#1}\small\normalsize} \spacingset{1}

\if0\blind
{
  \title{\bf A systematic assessment of Large Language Models for constructing two-level fractional factorial designs}
\author[1,$\dagger$]{Alan R. Vazquez \thanks{ORCID: \texttt{0000-0002-3658-0911}}}
\author[2]{Kilian M. Rother}
\author[1]{Marco V. Charles-Gonzalez}
\affil[1]{School of Engineering and Sciences, Tecnologico de Monterrey, Mexico}
\affil[2]{Faculty of Mechanical Engineering, University of Applied Sciences Kiel, Germany.}
\affil[$\dagger$]{Corresponding author. Email: {alanrvazquez@tec.mx};}
\affil[ ]{Contributing authors: {kilrother@googlemail.com}, {Marcocharlesgzz@gmail.com}}

  \maketitle
} \fi

\if1\blind
{
  \bigskip
  \bigskip
  \bigskip
  \begin{center}
    {\LARGE\bf A systematic assessment of Large Language Models for constructing two-level fractional factorial designs}
\end{center}
  \medskip
} \fi

\bigskip
\begin{abstract}
\noindent Two-level fractional factorial designs permit the study of multiple factors using a limited number of runs. Traditionally, these designs are obtained from catalogs available in standard textbooks or statistical software. However, modern Large Language Models (LLMs) can now produce two-level fractional factorial designs, but the quality of these designs has not been previously assessed. In this paper, we perform a systematic evaluation of two popular classes of LLMs, namely GPT and Gemini models, to construct two-level fractional factorial designs with 8, 16, and 32 runs, and 4 to 26 factors. To this end, we use prompting techniques to develop a high-quality set of design construction tasks for the LLMs. We compare the designs obtained by the LLMs with the best-known designs in terms of resolution and minimum aberration criteria. We show that the LLMs can effectively construct optimal 8-, 16-, and 32-run designs with up to eight factors. 
\end{abstract} 

\noindent%

\spacingset{1.4} 
\pagebreak
\section{Introduction}

Design of experiments (DoE) is a subfield of statistics that involves the design and analysis of cost-effective experiments to study complex processes. Typically, a process has several input factors that can be set deliberately and a response that measures the quality characteristic of the process' product. DoE methods, such as experimental plans and statistical models, are used to elucidate the true relationship between the response and the factors. 

Two-level fractional factorial designs \citep[][ch 5]{wu2011experiments} are commonly used experimental plans that can study many factors using an economical number of runs. In these designs, all factors are set at two levels, allowing the study of their main effects and interactions. We denote a two-level fractional factorial design as $2^{m-p}$, where $m$ is the number of factors and $p$ is the number of generated factors. The generated factors are obtained from the interactions of $b = m - p$ basic factors, whose $2^b$ level combinations are all in the design \citep{wu2011experiments}. Design $2^{m-p}$ is thus a $2^p$ fraction of the $2^m$ full factorial design, with a run size that is a power of two.

The DoE literature has complete catalogs of two-level fractional factorial designs. For instance, \cite{chen1993catalogue} provide a catalog of designs with 8, 16, 32, and 64 runs, and up to 32 factors. \cite{ryan2010minimum} provide a catalog of 128-run designs with up to 64 factors. The best designs in these catalogs are available in standard DoE textbooks such as \cite{wu2011experiments} and \cite{montgomery2017design}, or statistical software such as JMP, Minitab, and the FrF2 package \citep{gromping2014r} in R.

In recent years, generative artificial intelligence (GenAI) chatbots, such as ChatGPT \citep{chatgpt2025} and Gemini \citep{GoogleGemini}, have emerged as disruptive technology that has transformed the way we learn, write, access information, make decisions, and even conduct research. For example, GenAI chatbots have been used to solve optimization problems \citep{bertsimas2024robust}, prove mathematical theorems \citep{frieder2023mathematical}, teach statistics \citep{ellis2023new}, generate code for data analyses \citep{song2025performance}, and build search engine software for statistical quality control \citep{megahed2024introducing}. The success of GenAI chatbots is due to their use of Large Language Models \citep[LLMs; ][]{alammar2024hands} and the high computational power available these days \citep{top500}. Technically, an LLM is a deep neural network \citep{goodfellow2016deep} with billions of parameters that are estimated through a training process that uses an enormous amount of text data. These data are typically extracted from the Internet and may include books, blogs, research and news articles, social media, among others. The goal of the training process is to predict the next word in a sentence. To this end, the LLM uses a neural network architecture called \textit{transformer} \citep{vaswani2017attention}, which allows it to learn the context and relationships between words in the text during its training. 

In particular, ChatGPT and Gemini use the Generative Pre-trained Transformer \citep[GPT;][]{chatgpt2025} model and the Gemini model \citep{comanici2025gemini}, respectively, which are sophisticated, pre-trained transformer-based LLMs. To perform a specific task, their main input (as well as of other LLMs) is a user-specified text called \textit{prompt} \citep{schulhoff2024prompt}, which describes the question, instruction, or guidelines for it. At the time of writing this article, the versions of the LLMs used by default in ChatGPT's and Gemini's web interfaces are GPT-5.1 and Gemini 2.5 Flash, respectively. Remarkably, these LLMs can reason across audio, vision, and text in real time. Moreover, they can be accessed using application programming interfaces (APIs) for experimentation. 

In this paper, we evaluate the performance of the GPT-5.1 and Gemini 2.5 Flash models for constructing two-level fractional factorial designs. Our research goal is motivated by the following interaction between us and ChatGPT with the GPT-5.1 model available at \url{https://chatgpt.com/}. Prompt~\ref{ptp:introduction} shows the instruction we gave to ChatGPT. 
\begin{promptBox}{Simple prompt to construct a two-level design.} \label{ptp:introduction}
Construct a two-level fractional factorial design with four factors and eight runs.
\end{promptBox}

ChatGPT's output~\ref{out:introduction} shows a summary of the response of the chatbot to Prompt~\ref{ptp:introduction}. The symbol `$\vdots$' indicates that text has been omitted to streamline the output.
\begin{outputBox}{Summary of output to Prompt~\ref{ptp:introduction}.}\label{out:introduction}

A \textbf{two-level fractional factorial design with four factors (A, B, C, D) and eight runs} is a $2^{4-1}$ design: a half fraction of the full $2^4 = 16$ runs.

\textbf{Choose a generator}

A common choice is:

$$D=ABC$$

This produces a \textbf{Resolution IV} design.

\textbf{Construct the design}

Start with the full $2^3$ design for factors \textbf{A}, \textbf{B}, \textbf{C}, then set:

$$D=A \times B \times C$$

(where levels are coded as -1 and +1 or as - and +).

\textbf{Design Table (- = low, + = high)}

\begin{equation*}
    \begin{array}{cccccc}
\textbf{Run} & \textbf{A} & \textbf{B} & \textbf{C} & \textbf{D = ABC} & \\
1   & - & - & - & - & \\
2   & + & - & - & + & \\
3   & - & + & - & + & \\
4   & + & + & - & - & \\
5   & - & - & + & + & \\
6   & + & - & + & - & \\
7   & - & + & + & - & \\
8   & + & + & + & + & \\
    \end{array}
\end{equation*}
\begin{center}
$\vdots$
\end{center}
\end{outputBox}

ChatGPT's output~\ref{out:introduction} shows a $2^{4-1}$ design that matches the 4-factor 8-run design in Table 8.14 of \citet{montgomery2017design}. The four factors are labeled $A$, $B$, $C$, and $D$, where the first three are basic factors and the last is the generated factor obtained as $D = ABC$. That is, the column for $D$ is the element-wise product of the columns that accommodate $A$, $B$, and $C$, when their levels are coded as $-1$ and $1$. The design matrix is shown in the \textbf{Design Table} section. Its first column shows labels for the runs and the other columns show the factors, whose levels are labeled using the symbols `$-$' and `$+$'. This $2^{4-1}$ design can estimate the main effects and two-factor interactions of the four factors simultaneously. In fact, it is the best design in terms of the resolution and minimum aberration criteria introduced in Section~\ref{sec:criteria}. The Appendix shows a similar interaction between us and Gemini with the Gemini 2.5 Flash model, in which the same $2^{4-1}$ design was produced. 

Our interactions with ChatGPT and Gemini show that their default LLMs have the potential to construct two-level fractional factorial designs. Therefore, practitioners may feel tempted to construct these designs from scratch using these chatbots, bypassing statistical software and textbooks. To explore the capabilities and limitations of this approach, we conducted a systematic study to evaluate the performance of the GPT-5.1 and Gemini 2.5 Flash models on 36 tasks concerning the construction of two-level fractional factorial designs. The tasks involve designs with 8, 16, and 32 runs, and 4 to 26 factors. Each task has a prompt to instruct the LLM to construct a design. To this end, we develop a prompt template using prompting techniques such as role, context, chain of thought, and output format \citep{kojima2022large,white2023prompt,schulhoff2024prompt}. Our prompt template belongs to the class of \textit{zero-shot} prompts, introduced in Section~\ref{sec:instruction}, and is intended to test the general knowledge of LLMs on constructing two-level fractional factorial designs without examples.

We evaluate the computational performance of the LLMs using 10 replicates of each task. We show that the LLMs have the potential to construct optimal designs with up to 15 factors in terms of the criteria in Section~\ref{sec:criteria}. However, the Gemini 2.5 Flash model is better than the GPT-5.1 model because it has a high consistency in obtaining optimal 8-run designs, optimal 16-run designs with up to eight factors, and the optimal 32-run 6-factor design. For designs of other sizes, the performance of both LLMs deteriorates because they cannot consistently construct the optimal design or cannot construct a design at all. 

The remainder of the paper is organized as follows. In Section~\ref{sec:criteria}, we introduce the criteria for two-level fractional factorial designs. In Section~\ref{sec:instruction}, we develop our prompt template and provide the computational setup of the LLMs. In Section~\ref{sec:results}, we discuss the performance of the LLMs on our design construction tasks. In Section~\ref{sec:conclusion}, we end the paper with remarks and directions for future research. The R and Python codes to conduct our experiments are available on GitHub at \url{https://github.com/alanrvazquez/LLMforDOE}.


\section{Criteria for two-level fractional factorial designs} \label{sec:criteria}

We review resolution, minimum aberration, and minimum moment aberration \citep{wu2011experiments,xu2003minimum} to evaluate two-level fractional factorial designs. We illustrate these criteria using a $2^{7-3}$ design. Its basic factors are $A$, $B$, $C$, and $D$, and the generated factors are obtained as $E = ABC$, $F = ABD$, and $G = ACD$. The defining relation of the design is 
\begin{equation}\label{eq:DR_d1}
    I = ABCE = ABDF = ACDG = CDEF = BDEG = BCFG = AEFG.
\end{equation}
\noindent Recall that, in a defining relation, $I$ is the identity and the other elements are called words. For the benefit of the reader, Table~\ref{tab:D1} shows the $2^{7-3}$ design in full.

\begin{table}[htbp]
  \centering
  \caption{A two-level fractional factorial design with 16 runs and seven factors.} 
  \label{tab:D1}
  \begin{tabular}{rrrrrrrr}
    \toprule
Run	&	$A$	&	$B$	&	$C$	&	$D$	&	$E$	&	$F$	&	$G$	\\ \midrule 
1	&	$-1$	&	$-1$	&	$-1$	&	$-1$	&	$-1$	&	$-1$	&	$-1$	\\
2	&	$-1$	&	$-1$	&	$-1$	&	1	&	$-1$	&	1	&	1	\\
3	&	$-1$	&	$-1$	&	1	&	$-1$	&	1	&	$-1$	&	1	\\
4	&	$-1$	&	$-1$	&	1	&	1	&	1	&	1	&	$-1$	\\
5	&	$-1$	&	1	&	$-1$	&	$-1$	&	1	&	1	&	$-1$	\\
6	&	$-1$	&	1	&	$-1$	&	1	&	1	&	$-1$	&	1	\\
7	&	$-1$	&	1	&	1	&	$-1$	&	$-1$	&	1	&	1	\\
8	&	$-1$	&	1	&	1	&	1	&	$-1$	&	$-1$	&	$-1$	\\
9	&	1	&	$-1$	&	$-1$	&	$-1$	&	1	&	1	&	1	\\
10	&	1	&	$-1$	&	$-1$	&	1	&	1	&	$-1$	&	$-1$	\\
11	&	1	&	$-1$	&	1	&	$-1$	&	$-1$	&	1	&	$-1$	\\
12	&	1	&	$-1$	&	1	&	1	&	$-1$	&	$-1$	&	1	\\
13	&	1	&	1	&	$-1$	&	$-1$	&	$-1$	&	$-1$	&	1	\\
14	&	1	&	1	&	$-1$	&	1	&	$-1$	&	1	&	$-1$	\\
15	&	1	&	1	&	1	&	$-1$	&	1	&	$-1$	&	$-1$	\\
16	&	1	&	1	&	1	&	1	&	1	&	1	&	1	\\
    \bottomrule
  \end{tabular}
\end{table}

\subsection{Resolution and minimum aberration} \label{sec:resolution}

The resolution is the length of the shortest word in the defining relation \citep{wu2011experiments}. It provides a general summary of the properties of a $2^{m-p}$ design for studying main effects and two-factor interactions, which are the most important according to the effect hierarchy principle \citep[][ch 4]{wu2011experiments}. For instance, resolution-3 designs provide main effects that are not aliased with each other, but at least one main effect is aliased with one or more two-factor interactions. In contrast, resolution-4 designs provide main effects that are not aliased with each other nor with two-factor interactions. However, at least one pair of two-factor interactions is aliased in these designs. Designs with a resolution equal to five (or higher) are optimal for studying main effects and two-factor interactions because they avoid aliasing among these effects. So, when comparing designs in terms of their resolutions, the design with the higher resolution is preferred. From Equation~\eqref{eq:DR_d1}, we have that the $2^{7-3}$ design has a resolution of four.

The minimum aberration criterion allows us to differentiate between designs with the same resolution. It is based on the word length pattern \citep[WLP;][]{wu2011experiments}, which is a vector that collects the number of words of length 1 to $m$ in the defining relation. Specifically, this vector is $(A_1, A_2, \ldots, A_m)$ where $A_i$ is the number of words of length $i$ in the defining relation. Note that the resolution of a design is the smallest value of $i$ for which $A_i > 0$. From Equation~\eqref{eq:DR_d1}, we have that the WLP of the $2^{7-3}$ design is $(0, 0, 0, 7, 0, 0, 0)$. So, the defining relation of this design has seven words of length four. 

The minimum aberration criterion is to sequentially minimize the WLP from left to right. In this way, the criterion favors designs with less aliasing involving low-order effects of the factors than high-order ones. \citet{chen1993catalogue} show that the $2^{7-3}$ design in Table~\ref{tab:D1} has minimum aberration among all two-level fractional factorial designs of the same size.

\subsection{Minimum moment aberration} \label{sec:MMaberration}

For a large number of factors, comparing designs in terms of minimum aberration can become computationally expensive. This is because the computational complexity of the WLP is $O(n2^m)$, where $n$ is the design's run size. To overcome this issue, \cite{xu2003minimum} introduced minimum moment aberration to compare two-level designs. This criterion is equivalent to minimum aberration, but it uses the moment aberration pattern, which is a vector based on the similarity between the design's runs. 

Let $\mathbf{d}_i$ and $\mathbf{d}_j$ be the $i$-th and $j$-th run of a design. We consider $\delta(\mathbf{d}_i, \mathbf{d}_j)$ as a function that counts the number of entries in which $\mathbf{d}_i$ and $\mathbf{d}_j$ coincide. The $t$-th power moment of a two-level design is
\begin{equation*}
    K_t = \frac{\sum_{i=1}^{n-1} \sum_{j=i+1}^{n} \left[\delta(\mathbf{d}_i, \mathbf{d}_j)\right]^t}{n(n-1)/2}.
\end{equation*}

\noindent In statistical terms, $K_t$ is the $t$-th non-central moment of the empirical distribution of similarities between two runs of the design. The moment aberration pattern is $(K_1, K_2, \ldots, K_m)$, which has a computational complexity of $O(n^2 m^2)$ \citep{xu2003minimum}. This complexity is much lower than that of the WLP when $m$ is large. 

Similarly to minimum aberration, the minimum moment aberration criterion is to minimize the moment aberration pattern from left to right. The moment aberration pattern of the $2^{7-3}$ design in Table~\ref{tab:D1} is (3.27, 11.67, 42.47, 157.27, 591.27, 2251.67, 8666.47). Since this design has minimum aberration, it also has minimum moment aberration among all comparable designs.

In addition to its computational complexity, another advantage of minimum moment aberration is that it can evaluate two-level designs other than $2^{m-p}$ designs. For example, we can use it to evaluate two-level nonregular designs \citep[][ch 8]{wu2011experiments}, which are not constructed using basic and generated factors and so, do not have a defining relation. For this reason, we use minimum moment aberration in our numerical evaluations in Section~\ref{sec:results} to handle all possible types of two-level designs that LLMs may construct. 

In our evaluations, we also use the moment aberration pattern to compute the resolution of a two-level design. To this end, we use the lower bounds for the power moments of a two-level design of \cite{xu2003minimum}. Let $s$ be the smallest value of $i$ for which the $i$-th power moment of the design is equal to its lower bound $K'_i$. \cite{xu2003minimum} shows that the design projects into an equally replicated full $2^s$ factorial design in every subset of $s$ factors. In other words, the design has a \textit{strength} of $s$ \citep{hedayat2012orthogonal}. Note that, if the first power moment is higher than $K'_1$, $s=0$ and the design is not level-balanced. Following \cite{mee2017selecting}, we calculate the resolution of a two-level design as $s+1$. This calculation works for any two-level design and, for $2^{m-p}$ designs, it gives the length of the shortest word in the defining relation.

\section{Methodology} \label{sec:instruction}

Our methodology to construct two-level fractional factorial designs using LLMs has two main elements. The first one is a prompt template with precise instructions to construct designs, which we develop using prompting techniques \citep{white2023prompt,schulhoff2024prompt}. The second one is a computational setup for interacting with the LLMs, which involves APIs and dedicated Python packages. We present each of these elements separately. 

\subsection{Prompting techniques}

The input of an LLM is a prompt, the writing of which has a critical influence on the quality of the generated output \citep{schulhoff2024prompt}. To this end, there are several prompting techniques for crafting prompts, such as role, context, chain of thought, output format, and examples or \textit{shots} \citep{kojima2022large,white2023prompt,schulhoff2024prompt,sahoo2024systematic}. We use these techniques to develop our prompt template, except for the last one for three reasons. First, our template attempts to test the general knowledge of LLMs for constructing two-level fractional factorial designs from scratch using a single well-crafted instruction. Second, writing good examples of two-level fractional factorial designs and their construction requires profound DoE knowledge and expertise, which we do not assume our practitioner has. Third, adding examples to our prompt will significantly increase its length, making it more expensive to run in our APIs; see Section~\ref{sec:setupLLMs}. We comment on the benefits and challenges of prompts with examples, called \textit{few-shot} prompts in the prompting literature, in Section~\ref{sec:conclusion}. Since our prompt template does not include examples, it belongs to the class of \textit{zero-shot} prompts, which are relevant in practice \citep{schulhoff2024prompt,kojima2022large}. 

\subsubsection{Role}

The performance of an LLM for a given task can be improved by assigning its \textit{role} in the prompt \citep{wang2024rolellm}. This allows the LLM to focus on its expertise in the given role to provide the output. The technique of assigning a role (or \textit{persona}) to an LLM in the prompt is called role prompting \citep{wang2024rolellm}. \cite{arvidsson2023prompt}, \cite{kong2023better} and \cite{hadi2023large} discuss applications of role prompting in engineering, mathematics, and legal practice, respectively.  

To construct two-level fractional factorial designs, we assign the role of expert in DoE to the LLM. Specifically, we start our prompt template with the sentence: ``You are an expert in the subfield of statistics called design of experiments.'' In this role, we make explicit that DoE is a subfield of statistics, so that the LLM concentrates on the experimental design techniques discussed here. In other words, we expect the LLM to focus on two-level fractional factorial designs, resolution, and minimum aberration, which are standard topics in statistical DoE \citep{smucker2022,vazquez2025review}.

\subsubsection{Context}

The prompt given to an LLM must have the context needed to solve a given problem. This allows the LLM to generate a high-quality solution to the problem. If the prompt lacks context, the LLM might underachieve in the problem or generate a less informative output \citep{sahoo2024systematic,schulhoff2024prompt}. 

In our prompt template, we provide context on a design construction task in two ways. First, we state the goal in our prompt by adding the sentence: ``Your goal is to construct a two-level fractional factorial design with maximum resolution and minimum aberration.'' In this goal, we include the criteria for two-level designs in Section~\ref{sec:resolution}, which should be known by the LLM due to their role established in the previous section. So, we expect it to construct the best design in terms of these criteria. The goal does not involve minimum moment aberration because it is equivalent to minimum aberration; something that the LLM should know too because of its role.

Second, we state the desired number of factors, run size, and levels in the context given in our prompt template. Specifically, we add the following sentences to it: ``The number of factors is $m$ and the number of runs is $n$. The factors will be studied at two levels, which are coded as `$-$1' and `1'.'' These sentences state that the coded levels of the factors must be $-1$ and 1, which is standard in DoE. In the sentences, the symbols $m$ and $n$ must be replaced with the desired number of factors and runs, respectively.

\subsubsection{Zero-shot chain of thought}\label{sec:COT}

A chain of thought (CoT) is a series of intermediate natural language reasoning steps that can be added to a prompt to enhance the output of an LLM \citep{wei2022chain}. This technique allows the LLM to decompose the main problem into a series of smaller intermediate problems, each solved before giving the final output. \cite{wei2022chain} show that adding examples written using CoT to prompts improves the performance of transformer-based LLMs in arithmetic, symbolic reasoning, and common sense tasks.

\cite{kojima2022large} show that CoT can also improve the quality of zero-shot prompts. Specifically, they show that adding the sentence ``Let's think step by step,'' or a similar text to a zero-shot prompt, generally enhances the performance of transformer-based LLMs in similar tasks as \cite{wei2022chain}. \cite{kojima2022large} refer to the technique of adding this sentence (or one of its variants) to a zero-shot prompt as Zero-shot-CoT. Attractive features of this technique include its simplicity, task independence, and the fact that it avoids the need for good CoT-based examples, which require significant expertise and careful writing. 

We use Zero-shot-CoT in our prompt template by including the sentence ``You will think step by step about how to construct the design.'' In this way, we expect to leverage the reasoning capabilities of the LLMs used here to produce attractive designs.

\subsubsection{Output format}\label{sec:output_template}

We can set a format for an LLM to tailor the structure of their output to our needs \citep{white2023prompt}. For example, we can provide constraints on the length and format of its output in the prompt. An output format is particularly useful for our problem because the LLMs tend to provide two-level designs with additional text to explain its properties and construction; see ChatGPT's output~1 and Gemini's output~A1 in the Appendix. However, for our numerical experiments, we only need a table with the design in a specific format. That is, the table must have $m$ columns and $n$ rows with entries equal to $-1$ or 1. The table must also have a header row with the labels of the factor columns and an additional column to label the runs. 

To achieve the desired format for the design table, we develop an output format with two elements that are included in our prompt template. The first one involves two sentences to avoid the explanations of the LLM and produce a design table only. These sentences are: ``However, you will only generate a table containing the design. You will not generate any text explaining the table or your answer.'' 

The second element of our output format allows us to streamline the LLMs output because it sets the format for the table. Specifically, it instructs the LLM to generate the table using a comma-separated values (CSV) format, which is common for storing tables, matrices, and arrays. In this format, every entry in the table is separated by a comma. We specify the end of a row using two backslash symbols (\textbackslash\textbackslash). To set our second format element, we add the following sentences to our prompt: ``The table must be in a comma-separated values (CSV) format. Specifically, the values in the table must be separated by `,’ and each row must end with `\textbackslash\textbackslash’. In the table, the first row will be used as a header row to label the factors using the letters of the English alphabet starting with `A’. The first column will be called ``Run'' and used to count the number of runs starting at `1’. Each design cell (excluding the header and Run columns) must contain either `-1’ or `1’.'' We instruct the LLMs to label factors using the letters of the English alphabet following \cite{montgomery2017design}, ChatGPT's output~\ref{out:introduction}, and Gemini's output~A1. Moreover, the maximum number of factors in our experiments in Section~\ref{sec:results} is 26, which is the number of letters in this alphabet. 

Our output format results in a design table suitable for evaluation using the criteria in Section~\ref{sec:criteria} and our Python code. To our knowledge, this is the first output format used to obtain a two-level fractional factorial design table from an LLM.

\subsubsection{Prompt template}

Using the previous prompting techniques, we develop the prompt template in Prompt~2 to construct two-level fractional factorial designs using LLMs. The template has a final instruction to restate the goal of the task. That is, Prompt~2 ends with the sentence ``Construct the two-level fractional factorial design with $m$ factors and $n$ runs that has maximum resolution and minimum aberration.'' Recall that the symbols $m$ and $n$ must be replaced with the desired number of factors and runs, respectively. Due to its use of the Zero-shot-CoT technique, we refer to Prompt~2 as a Zero-shot-CoT prompt. 

\begin{promptBox}{Prompt template to construct a two-level design.} \label{ptp:expert}
You are an expert in the subfield of statistics called design of experiments. Your goal is to construct a two-level fractional factorial design with maximum resolution and minimum aberration. The number of factors is $m$ and the number of runs is $n$. The factors have two levels coded as `-1' and `1'. You will think step by step about how to construct the design. However, you will only generate a table containing the design. You will not generate any text explaining the table or your answer. The table must be in a comma-separated values (CSV) format. Specifically, the values in the table must be separated by `,’ and each row must end with `\textbackslash\textbackslash’. In the table, the first row will be used as a header row to label the factors using the letters of the English alphabet starting with `A’. The first column will be called ``Run'' and used to count the number of runs starting at `1’. Each design cell (excluding the header and Run columns) must contain either `-1’ or `1’. Construct the two-level fractional factorial design with $n$ runs and $m$ factors that has maximum resolution and minimum aberration.
\end{promptBox}

Compared to Prompt~\ref{ptp:introduction}, Prompt~2 provides richer instructions to an LLM because it defines its role, sets the goal and context of the task, requests to complete the task step by step, and provides a template for the output. Therefore, Prompt~2 has the potential to produce two-level fractional factorial designs with good statistical properties.


\subsection{\revb{Setup for the LLMs}} \label{sec:setupLLMs}

To construct designs with Prompt~2, we use the GPT-5.1 and Gemini 2.5 Flash models, which are the default LLMs in ChatGPT's and Gemini's web interfaces, respectively, at the time of writing this article. Technically, these LLMs are \textbf{gpt-5.1-chat-latest} \citep{openai_gpt5_1_chat_latest} and \textbf{gemini-2.5-flash} \citep{google_gemini_2_5_flash}. We run them using the Python packages called \verb!openai! \citep{openai_python} and \verb!google-genai! \citep{google-genai}, and API keys from OpenAI and Google. An API key is a unique code that can be used to access and communicate with an API, which, in turn, provides access to an LLM. The \verb!openai! and \verb!google-genai! packages provide user-friendly Python functions to interact with the APIs.

Compared to ChatGPT's and Gemini's web interfaces, our setup allows us to perform multiple design construction tasks with Prompt~2 and the LLMs automatically. It also allows us to set up a separate conversation involving Prompt~2 and an LLM for each design construction task. In this way, we ensure the behavior of our practitioner across all tasks. That is, a user whose first query for the LLM is Prompt~2 and for whom there is no prior interaction history with the LLM. Although we can create independent conversations with the LLMs in ChatGPT's and Gemini's web interfaces, we must do this manually, which makes it difficult to perform all of our design construction tasks in the next section.

The Python packages and APIs enable more tuning parameters of the LLMs than ChatGPT's and Gemini's web interfaces. For example, through the \verb!google-genai! package, we can set the parameters called \textit{temperature}, \textit{top-k}, and \textit{top-p} of \textbf{gemini-2.5-flash}, which control the randomness and diversity of its output to multiple runs of the same prompt \citep{vertex-ai_content_generation_parameters}. We can also set the parameter called \textit{thinkingBudget}, which defines the strategy for reasoning of this LLM. The \verb!openai! package provides access to similar tuning parameters of \textbf{gpt-5.1-chat-latest}. In our experiments, we use the default values of all tuning parameters of \textbf{gpt-5.1-chat-latest} and \textbf{gemini-2.5-flash}, except for the parameter that controls their reasoning. For \textbf{gpt-5.1-chat-latest}, we set the \textit{reasoning-effort} parameter to ``medium.'' For \textbf{gemini-2.5-flash}, we set \textit{thinkingBudget} to ``-1'' and use the dynamic thinking strategy, which lets the LLM decide when and how much to think based on the complexity of the request. 

Running an LLM using an API key has associated costs that depend on the number of \textit{tokens} involved in the input and output of the model. Essentially, tokens are the data units of an LLM.  They are constructed from text by the LLM using a process called \textit{tokenization}, which considers language, punctuation, and context, among other features. For input and output text, \textbf{gpt-5.1-chat-latest} charges \$1.5 and \$10 for one million tokens, respectively \citep{openai_gpt5_1_chat_latest}. For \textbf{gemini-2.5-flash}, these costs are \$0.3 and \$2.5 \citep{google_gemini_api_pricing_2025}. Therefore, working with long prompts and outputs is more expensive than with short ones. In this regard, a benefit of our Zero-shot-CoT prompt template is that it allows us to limit the costs of our numerical experiments, because they involve many executions of Prompt~2. 


\section{Numerical results} \label{sec:results}

We evaluate the performance of the GPT-5.1 and Gemini 2.5 Flash models for design construction tasks involving two-level fractional factorial designs with 8, 16, and 32 runs. For these run sizes, the number of factors in our tasks ranges from 4 to 7, 5 to 15, and 6 to 26, respectively. The number of tasks thus is 36. For each task, we construct 10 designs from 10 independent executions of an LLM with Prompt~2 as input. Therefore, we construct a total of $10 \times 36 = 360$ designs using each LLM. The order in which we perform the tasks with each LLM is random. 

\revb{We classify the designs obtained from an LLM as compliant or non-compliant. A design is compliant if the resulting design table has complete entries and the requested number of factors, levels, and runs in the task. Otherwise, the design is non-compliant. In some of our experiments, the compliant designs did not strictly follow the output format in Prompt~2. This was mainly because they had extra commas or missed a backslash symbol at the end of a row of the table. In these cases, we manually corrected their format so that they could be evaluated using our Python code.} 

We evaluate the compliant designs in terms of resolution and minimum moment aberration calculated as in Section~\ref{sec:MMaberration}. In Section~\ref{sec:GPTresults}, we show the numerical results of the GPT-5.1 model for the 8-, 16-, and 32-run design tasks separately. In Section~\ref{sec:Geminiresults}, we discuss the performance of the Gemini 2.5 Flash model on these tasks. 

\subsection{Designs constructed using the GPT-5.1 model}\label{sec:GPTresults}

\subsubsection{Tasks with eight runs}
Table~\ref{tab:8run_expert} shows the performance of the GPT-5.1 model on the design construction tasks with eight runs. Specifically, the table shows the task label and the maximum, minimum, and median resolutions of the designs obtained by this LLM. For each task, the 10 designs constructed by the LLM were compliant.

\begin{table}[htbp]
  \centering
  \caption{Performance of the GPT-5.1 model on 8-run design tasks.}
  \label{tab:8run_expert}
  \begin{tabular}{ccccc}
    \toprule
            &  & \multicolumn{3}{c}{Resolution} \\ \cmidrule{3-5}
Task &    Factors & Min. & Median & Max.  \\ \midrule
1 &    4        & 4 & 4 & 4  \\
2 &    5        & 2 & 3 & 3 \\
3 &    6        & 2 & 2.5 & 3 \\
4 &    7        & 3 & 3 & 3 \\
    \bottomrule
  \end{tabular}
\end{table}

Table~\ref{tab:8run_expert} shows that the resolution of the designs with four and seven factors is four and three, respectively. For five and six factors, the resolution is two or three. In contrast, all two-level fractional factorial designs with eight runs and four to seven factors in  \cite{wu2011experiments} have a resolution of at least three. The same is true for the 16- and 32-run fractional factorial designs, which we discuss later. 

For each number of factors, the resolution of the best design obtained by the GPT-5.1 model equals that of the minimum aberration design in \cite{wu2011experiments}. To further compare these designs, we computed their moment aberration patterns. Table~\ref{tab:MA_designs_8run} shows the patterns of the best designs obtained by the LLM for each task. The best designs have the same moment aberration pattern as the minimum aberration designs. Therefore, the GPT-5.1 model constructed the optimal designs for the 8-run design tasks. Regarding the consistency in obtaining them, it found the optimal 4-, 5-, 6-, and 7-factor designs in 10, 8, 5, and 10, respectively, of the 10 independent executions of Prompt~2; see Table~\ref{tab:MA_designs_8run}.

\begin{table}[htbp]
  \centering
  \caption{Moment aberration patterns obtained by the GPT-5.1 model for the 8-run design tasks.}
  \label{tab:MA_designs_8run}
  \begin{tabular}{ccclc}
    \toprule
Task & Factors & Resolution & Moment aberration pattern & Frequency \\ \midrule
1 & 4 & 4 & (1.7, 3.4, 6.9, 13.7) & 10       \\
2 & 5 & 3 & (2.1, 5.0, 12.4, 32.4, 87.9) & 8       \\
3 & 6  & 3 &  (2.6, 6.9, 18.9, 53.1, 152.6, 444.0)  & 5     \\
4 & 7  & 3 &  (3.0, 9.0, 27.0, 81.0, 243.0, 729.0, 2187.0) & 10\\
    \bottomrule
  \end{tabular}  
\end{table}

\subsubsection{Tasks with 16 runs}

Table~\ref{tab:16run_expert} summarizes the resolutions of the designs obtained by the GPT-5.1 model for the 16-run design tasks. The table also shows the number of compliant designs that were generated for each task. The LLM obtained 10 compliant designs for each task, except for the tasks with 12 to 15 factors. For these tasks, it only obtained eight or nine compliant designs. A close inspection to the non-compliant designs revealed that they had missing entries or were infeasible for the LLM. For the latter case, the LLM output a sentence similar to: ``I'm sorry, but I cannot generate a valid design that meets your requirements.''


Table~\ref{tab:16run_expert} shows that the median resolutions of the 16-run designs are higher than or equal to three, except for the designs with 11, 14, and 15 factors. For these designs, the median resolution was 1.5, 2 or 2.5. In general, the compliant designs of resolution one and two were not $2^{m-p}$ designs. This was also the case for the 32-run designs in the next section, and the designs obtained by the Gemini 2.5 Flash model in Section~\ref{sec:Geminiresults}.

\begin{table}[htbp]
  \centering
  \caption{Performance of the GPT-5.1 model on 16-run design tasks.}
  \label{tab:16run_expert}
  \begin{tabular}{ccccc}
    \toprule
            & \multicolumn{3}{c}{Resolution} & \\ \cmidrule{2-4}
Task/Factors & Min. & Median & Max. & \# Compliant\\ \midrule
5        & 1 & 5 & 5 & 10\\
6        & 1 & 4 & 4  & 10\\
7        & 1 & 4 & 4 & 10\\
8        & 1 & 4 & 4 & 10\\
9        & 1 & 3 & 3 & 10\\
10        & 1  & 3 & 3 & 10\\
11        & 1 & 2.5 & 3 & 10\\
12        & 1 & 3 & 3 & 9\\ 
13        & 1 & 3 & 3 & 8 \\ 
14        & 1 & 2 & 3 & 8 \\ 
15        & 1 & 1.5 & 3 & 8 \\   
    \bottomrule
  \end{tabular}
\end{table}

For all numbers of factors, there were multiple compliant designs with the same resolution as the corresponding minimum aberration designs \citep{wu2011experiments}. Table~\ref{tab:MA_16run_designs} shows the frequencies of the moment aberration patterns of these designs for each task. To enhance its layout, the table shows the first four elements of the moment aberration pattern. For each number of factors, the GPT-5.1 model obtained the minimum aberration design at least twice, except for 12 factors for which it did not find the minimum aberration design. Interestingly, for tasks with five to eight factors, this LLM obtained the minimum aberration design in at least eight repetitions of these tasks. Therefore, the GPT-5.1 model performs well in constructing the optimal two-level fractional factorial designs with 16 runs and up to eight factors.

\begin{table}[htbp]
  \centering
  \caption{Moment aberration patterns obtained by the GPT-5.1 model for the 16-run design tasks. $^\star$: Design has minimum moment aberration.}
  \label{tab:MA_16run_designs}
  \begin{tabular}{cclc}
    \toprule
Task/Factors & Resolution & Moment aberration pattern & Frequency \\ \midrule
5	&	5	& (2.3, 6.3, 18.3, 54.3)$^\star$ 	&	9	\\
6	&	4	& (2.8, 8.8, 28.8, 97.6)$^\star$ 	&	8	\\
7	&	4	& (3.3, 11.7, 42.5, 157.3)$^\star$ 	&	8	\\
8	&	4	& (3.7, 14.9, 59.7, 238.9)$^\star$ 	&	8	\\
9	&	3	& (4.2, 18.6, 84.2, 386.6)$^\star$ 	&	3	\\
	&		& (4.2, 18.6, 86.6, 421.8) 	&	2	\\
	&		& (4.2, 18.6, 87.4, 437.8) 	&	1	\\
10	&	3	& (4.7, 22.7, 113.1, 575.5)$^\star$	&	2	\\
	&		& (4.7, 22.7, 113.9, 588.3) 	&	1	\\
	&		& (4.7, 22.7, 114.7, 602.7) 	&	3	\\
	&		& (4.7, 22.7, 114.7, 604.3) 	&	1	\\
11	&	3	& (5.1, 27.1, 146.7, 807.9)$^\star$ 	&	2	\\
	&		& (5.1, 27.1, 147.5, 823.9) 	&	2	\\
	&		& (5.1, 27.1, 147.5, 825.5) 	&	1	\\
12	&	3	& (5.6, 32, 186.4, 1105.6) 	&	5	\\
13	&	3	& (6.1, 37.3, 231.7, 1456.5)$^\star$ 	&	5	\\
14	&	3	& (6.5, 42.9, 283.7, 1885.3)$^\star$ 	&	3	\\
15	&	3	& (7.0, 49, 343, 2401.0)$^\star$ 	&	3	\\
    \bottomrule
  \end{tabular}  
\end{table}

\subsubsection{Tasks with 32 runs}

Table~\ref{tab:32run_expert} shows the resolution and number of compliant designs obtained by the GPT-5.1 model for the 32-run design tasks. For 11, 13, 16, and 19 to 26 factors, this LLM did not find a compliant design with a resolution higher than two. For this reason, these numbers of factors are omitted in the table. Table~\ref{tab:32run_expert} shows that there were few compliant designs for tasks with more than eight factors. Similar to the 16-run design tasks, the 32-run noncompliant designs had missing entries or were infeasible for the LLM. 

\begin{table}[htbp]
  \centering
  \caption{Performance of the GPT-5.1 model on 32-run design tasks.}
  \label{tab:32run_expert}
  \begin{tabular}{cccccc}
    \toprule
            & & \multicolumn{3}{c}{Resolution} & \\ \cmidrule{3-5}
Task & Factors & Min. & Median & Max. & \# Compliant\\ \midrule
16  &   6   & 1 & 6 & 6 & 10\\
17  &   7   & 1 & 4 & 4  & 9\\
18  &    8  & 1 & 3.5 & 4 & 8\\
19  &  9    & 1 & 1 & 4 & 3\\
20  &   10   & 1 & 2 & 3 & 2\\
22  &  12    & 2 & 3 & 4 & 2\\
24  &    14  & 1 & 2 & 3 & 4 \\ 
25  &   15   & 1 & 2.5 & 3 & 4 \\ 
27  &    17  & 1 & 2 & 3 & 2 \\ 
28  &    18  & 2 & 2.5 & 3 & 2 \\ 
    \bottomrule
  \end{tabular}
\end{table}

For 6-9, 12, 17, and 18 factors, the maximum resolution in Table~\ref{tab:32run_expert} matches that of the minimum aberration designs \citep{wu2011experiments}. For these numbers of factors, Table~\ref{tab:MA_32run_designs} shows the first six elements of the moment aberration pattern of the designs with the highest resolution obtained by the GPT-5.1 model. For 6-9 and 12 factors, the LLM found the minimum aberration design at least once. This is because the best moment aberration pattern is equal to that of the minimum aberration design for these cases. For 17 and 18 factors, the best moment aberration patterns are worse than those of the minimum aberration designs. Overall, we conclude that the GPT-5.1 model is moderately effective for constructing the 32-run 6-factor minimum aberration design, but ineffective for generating 32-run minimum aberration designs with more factors.

\begin{table}[htbp]
  \centering
  \caption{Moment aberration patterns obtained by the GPT-5.1 model for selected 32-run design tasks. $^\star$: Design has minimum moment aberration.}
  \label{tab:MA_32run_designs}
  \begin{tabular}{ccclc}
    \toprule
Task & Factors & Resolution & Moment aberration pattern & Frequency \\ \midrule
16	&	6	&	6	& (2.9, 9.7, 34.8, 131.6, 511.0, 2012.9)$^\star$ &	6	\\
17	&	7	&	4	& (3.4, 12.9, 52.2, 220.4, 959.5, 4278.7)$^\star$ &	1	\\
	&		&		& (3.4, 12.9, 52.2, 221.9, 978.9, 4437.4) &	1	\\
	&		&		& (3.4, 12.9, 52.2, 223.5, 1006.0, 4735.5) &	4	\\
18	&	8	&	4	& (3.9, 16.5, 74.3, 346.3, 1656.8, 8099.1)$^\star$ &	1	\\
	&		&		& (3.9, 16.5, 74.3, 349.4, 1703.2, 8555.9) &	1	\\
	&		&		& (3.9, 16.5, 74.3, 352.5, 1765.2, 9337.8) &	2	\\
19	&	9	&	4	& (4.4, 20.6, 101.9, 517.6, 2683.1, 14141.9)$^\star$ &	1	\\
22	&	12	&	4	& (5.8, 35.6, 223.0, 1423.0, 9243.9, 61042.1)$^\star$ &	1	\\
27	&	17	&	3	& (8.2, 69.6, 608.2, 5485.9, 51108.9, 491295.5) &	1	\\
28	&	18	&	3	& (8.7, 77.8, 714.6, 6758.7, 65883.9, 661601.0) &	1	\\
    \bottomrule
  \end{tabular}  
\end{table}

\subsection{Designs constructed using the Gemini 2.5 Flash model} \label{sec:Geminiresults}

\revb{We performed the design construction tasks with 8, 16, and 32 runs using the Gemini 2.5 Flash model. Remarkably, this LLM excelled in the 8-run design tasks because it obtained the minimum aberration designs for all factors in all independent executions of Prompt~2. This LLM also obtained 10 compliant designs for each 16-run design task. Table~\ref{tab:16run_gemini} summarizes the resolutions of these designs. 

The Gemini 2.5 Flash model constructed 16-run designs with the same resolution as the minimum aberration designs. Table~\ref{tab:MA_16run_designs_gemini} shows the moment aberration pattern of the 16-run designs constructed by this LLM. The table is akin to Table~\ref{tab:MA_16run_designs} for the GPT-5.1 model, and it concerns the designs with the highest resolution obtained for each task. Except for 11 and 12 factors, the Gemini 2.5 Flash model obtained the 16-run minimum aberration designs at least twice; see Table~\ref{tab:MA_16run_designs_gemini}.}

\begin{table}[htbp]
  \centering
  \caption{Performance of the Gemini 2.5 Flash model on the 16-run design tasks.}
  \label{tab:16run_gemini}
  \begin{tabular}{cccc}
    \toprule
            & \multicolumn{3}{c}{Resolution} \\ \cmidrule{2-4}
Task/Factors & Min. & Median & Max. \\ \midrule 
5        & 3 & 5 & 5 \\
6        & 1 & 4 & 4 \\
7        & 1 & 4 & 4 \\ 
8        & 1 & 4 & 4 \\ 
9        & 1 & 3 & 3 \\ 
10        & 2  & 3 & 3 \\ 
11        & 1 & 2 & 3 \\ 
12        & 1 & 2.5 & 3 \\ 
13        & 1 & 1.5 & 3  \\ 
14        & 1 & 1 & 3  \\ 
15        & 1 & 1 & 3  \\ 
    \bottomrule
  \end{tabular}
\end{table}

\begin{table}[htbp]
  \centering
  \caption{Moment aberration patterns of obtained by the Gemini 2.5 Flash model for selected 16-run design tasks. $^\star$: Design has minimum moment aberration.}
  \label{tab:MA_16run_designs_gemini}
  \begin{tabular}{cclc}
    \toprule
Task/Factors & Resolution & Moment aberration pattern & Frequency \\ \midrule
5	&	5	& (2.3, 6.3, 18.3, 54.3)$^\star$ 	&	8	\\
6	&	4	& (2.8, 8.8, 28.8, 97.6)$^\star$ 	&	9	\\
7	&	4	& (3.3, 11.7, 42.5, 157.3)$^\star$ 	&	9	\\
8	&	4	& (3.7, 14.9, 59.7, 238.9)$^\star$ 	&	8	\\
9	&	3	& (4.2, 18.6, 84.2, 386.6)$^\star$ 	&	2	\\
	&		& (4.2, 18.6, 86.6, 421.8) 	&	4	\\
10	&	3	& (4.7, 22.7, 113.1, 575.5)$^\star$ 	&	3	\\
	&		& (4.7, 22.7, 113.9, 588.3) 	&	1	\\
	&		& (4.7, 22.7, 114.7, 602.7) 	&	5	\\
11	&	3	& (5.1, 27.1, 147.5, 823.9) 	&	3	\\
12	&	3	& (5.6, 32, 186.4, 1105.6) 	&	5	\\
13	&	3	& (6.1, 37.3, 231.7, 1456.5)$^\star$ 	&	4	\\
14	&	3	& (6.5, 42.9, 283.7, 1885.3)$^\star$ 	&	3	\\
15	&	3	& (7.0, 49.0, 343.0, 2401.0)$^\star$ 	&	3	\\
    \bottomrule
  \end{tabular}  
\end{table}

Regarding the 32-run design tasks, the LLM constructed 10 compliant designs for each number of factors, except for 17, 18, and 20 to 26 factors. For these cases, the LLM generated several non-compliant designs with missing entries or other issues. For tasks with 6-9, 10, and 14 factors, the compliant designs had a resolution higher than or equal to three. Table~\ref{tab:32run_gemini} shows the distribution of the resolution  for these cases. For six to nine factors, the maximum resolutions in this table equal those of comparable 32-run minimum aberration designs. Table~\ref{tab:MA_32run_designs_gemini}, which is akin to Table~\ref{tab:MA_32run_designs}, shows that the Gemini 2.5 Flash model obtained the minimum aberration designs with six to eight factors at least once. Remarkably, for six factors, the LLM constructed the 32-run minimum aberration design in the 10 independent executions of the Prompt~2.

\begin{table}[htbp]
  \centering
  \caption{Performance of the Gemini 2.5 Flash model on 32-run design tasks.}
  \label{tab:32run_gemini} 
  \begin{tabular}{ccccc}
    \toprule
            & & \multicolumn{3}{c}{Resolution} \\ \cmidrule{3-5}
Task & Factors & Min. & Median & Max. \\ \midrule
16  &  6   & 6 & 6 & 6 \\
17  &  7   & 2 & 3 & 4 \\
18  &  8  & 2 & 3 & 4 \\
19  &  9    & 1 & 3 & 4 \\
20  &  10   & 1 & 2 & 3 \\ 
24  &  14  & 1 & 2 & 3  \\ 
    \bottomrule
  \end{tabular}
\end{table}

\begin{table}[htbp]
  \centering
  \caption{Moment aberration patterns obtained by the Gemini 2.5 Flash model for selected 32-run design tasks. $^\star$: Design has minimum moment aberration.}
  \label{tab:MA_32run_designs_gemini}
  \begin{tabular}{ccclc}
    \toprule
Task & Factors & Resolution & Moment aberration pattern & Frequency \\ \midrule
16	&	6	&	6	& (2.9, 9.7, 34.8, 131.6, 511.0, 2012.9)$^\star$ &	10	\\
17	&	7	&	4	& (3.4, 12.9, 52.2, 220.4, 959.5, 4278.7)$^\star$ &	2	\\
	&		&		& (3.4, 12.9, 52.2, 221.9, 978.9, 4437.4) &	1	\\
18	&	8	&	4	& (3.9, 16.5, 74.3, 346.3, 1656.8, 8099.1)$^\star$ &	1	\\
	&		&		& (3.9, 16.5, 74.3, 349.4, 1703.2, 8555.9) &	1	\\
19	&	9	&	4	& (4.4, 20.6, 101.9, 522.3, 2756.6, 14896.7) &	1	\\
    \bottomrule
  \end{tabular}  
\end{table}

\revb{Compared to the GPT-5.1 model, the Gemini 2.5 Flash model had a success rate of 100\% in constructing minimum aberration designs with eight runs and four to seven factors, and with 32 runs and six factors. Moreover, its success rate for constructing 16-run minimum aberration designs with five to eight factors was at least 80\%. However, the GPT-5.1 and Gemini 2.5 Flash models were not successful in constructing minimum aberration designs of other sizes, since their success rate was 50\% or less.}   




\section{Conclusion} \label{sec:conclusion}

We assessed the quality of two-level fractional factorial designs constructed by the GPT-5.1 and Gemini 2.5 Flash models using Zero-shot-CoT prompts. To this end, we developed a prompt template to tackle design construction tasks with 8, 16, and 32 runs, and 4 to 26 factors. Using the template as input, we generated 10 designs for each task using 10 independent executions of the LLMs. We compared the designs obtained with the optimal designs in terms of resolution, minimum aberration, and minimum moment aberration. We showed that the LLMs constructed all 8- and 16-run minimum aberration designs at least twice, except for the design with 16 runs and 12 factors. They also generated  32-run minimum aberration designs with six to nine and 12 factors at least once. 

Based on our results, we recommend Gemini with its Gemini 2.5 Flash model to construct 8-run designs with four to seven factors, and a 32-run design with six factors using Prompt~2. We also recommend Gemini and this prompt to generate 16-run designs with five to eight factors, but warn practitioners that there is a small chance that this chatbot will produce poor designs. On the upside, these numbers of factors are the most common in practice according to the systematic surveys of real factorial experiments conducted by \cite{li2006regularities} and \cite{ockuly2017response}. On the downside, the Gemini 2.5 Flash model is infective in constructing attractive designs with more factors consistently using Prompt~2. For these cases, we advise practitioners to use DoE textbooks \citep{wu2011experiments,montgomery2017design}, JMP, Minitab, or the FrF2 package in R.

There are two potential ways to improve the performance of the GPT-5.1 and Gemini 2.5 Flash models on our design construction tasks. One way is to use a \textit{few-shot} prompt template with CoT-based examples instead of our Zero-shot-CoT prompt template. Despite the improvement in performance of LLMs given by Zero-shot-CoT prompts, \cite{kojima2022large} show that these prompts do not outperform few-shot prompts with CoT-based examples in general. However, developing these examples is challenging because their writing must have all the steps needed to reach the solution of a task's instance. Moreover, the writing style of these steps influences the performance of LLMs \citep{kojima2022large}. In our setting, developing CoT-based examples would need a careful selection of two-level fractional factorial designs, as they can have different resolutions and WLPs. Moreover, each example would need a comprehensive step-by-step explanation of their construction. Such an explanation should include the selection of basic and generated factors, the elaboration of the defining relation and WLP, and the evaluation in terms of resolution and minimum aberration. For all these reasons, we leave the use of few-shot prompts with CoT-based examples for constructing designs as a topic for future research.  

Another way to improve the performance of the GPT-5.1 and Gemini 2.5 Flash models is using Retrieval-Augmented Generation \citep[RAG;][]{fan2024survey}. RAG is a methodology for enhancing the performance of pre-trained LLMs using domain-specific sources provided by the user. In our context, these sources can be textbooks and research articles on DoE that are open-source \citep{Jones-Farmer19102024}. RAG has three main steps. First, it stores the domain-specific sources in a database using encoding methods. Second, for an input prompt, it retrieves the elements from the database that are most similar to the prompt. Third, it informs the LLM about these elements by, for example, using them context in the prompt. In this way, RAG helps the LLM generate domain-specific output for a given task. We refer the reader to \cite{megahed2024introducing} for an application of RAG in statistical quality control, and leave its application for constructing two-level fractional factorial designs as another topic for future research. 

Finally, we studied the default versions of the GPT and Gemini models used by ChatGPT's and Gemini's web interfaces at the time of writing this article (December, 2025). Due to the rapid development of research and applications of artificial intelligence, we expect the new versions of these LLMs---or even of other LLMs not studied here, such as the LLaMA model \citep{touvron2023llama}---to become more capable in domain-specific tasks involving Zero-shot-CoT prompts. Our set of 36 design construction tasks thus enriches the catalog of benchmarks used to assess the gains in performance of these versions and LLMs \citep{ni2025survey}. We hope to see future studies showcasing the performance of these and other GenAI technologies in our set.  



\begin{center}
{\large\bf ACKNOWLEDGMENTS}
\end{center}
The authors acknowledge that ChatGPT and Gemini were used for basic queries about LLMs and APIs, to construct two-level fractional factorial designs, and for language assistance in some sentences through Writefull (\url{https://www.writefull.com/}). They were not used to generate or explore ideas, classify literature, or provide coding assistance. The research of Vazquez is supported by the Challenge-Based Research Fund of Tecnologico de Monterrey under the project CI-EIC-HLT-D-58.  

\begin{center}
{\large\bf DISCLOSURE OF INTEREST}
\end{center}
The authors report that there are no competing interests to declare.

\begin{center}
{\large\bf DECLARATION OF FUNDING}
\end{center}
No funding was received.

\begin{center}
{\large\bf DATA AVAILABILITY STATEMENT}
\end{center}
All data and code related to this paper are available on GitHub at \url{https://github.com/alanrvazquez/LLMforDOE}.

\bibliographystyle{apalike} 
\bibliography{LLMdoe} 

\appendix
\setcounter{table}{0}
\renewcommand{\thetable}{A\arabic{table}}

\begin{center}
{\large\bf APPENDIX}
\end{center}

Gemini's output~A1 shows a summary of an interaction between us and this GenAI chatbot available at \url{https://gemini.google.com/app}. The input to Gemini was Prompt~\ref{ptp:introduction} and the LLM used was the Gemini 2.5 Flash model. Gemini produced the best $2^{4-1}$ design in terms of the resolution and minimum aberration criteria. 

\begin{outputBoxGemini}{Summary of output to Prompt~\ref{ptp:introduction}.}\label{out:introduction_gemini}

A two-level fractional factorial design with four factors ($k=4$) and eight runs ($n=8$) is a $2^{4-1}$ \textbf{design}. The $2^{4-1}$ design is a \textbf{half-fraction} of the full $2^4$ factorial design, resulting in a \textbf{Resolution IV} design.

\textbf{1. Defining the Design}

The full $2^4$ design has $2^4 = 16$ runs. A half-fraction means we are performing $16/2 = 8$ runs.

\begin{itemize}
\item Factors: Let the four factors be $\mathbf{A}$, $\mathbf{B}$, $\mathbf{C}$, and $\mathbf{D}$.
\item Levels: The two levels for each factor are denoted by -1 (low) and +1 (high).
\item Defining Relation/Generator: To create the half-fraction, we need one generator or designate that relates the last factor, $\mathbf{D}$, to the interactions of the previous factors $\mathbf{A}$, $\mathbf{B}$, and $\mathbf{C}$. The choice of the generator determines the confounding pattern (aliasing) of the design.
\end{itemize}

\begin{center}
$\vdots$
\end{center}

\textbf{3. Final Fractional Factorial Design}

The resulting $2^{4-1}$ fractional factorial design with $D=ABC$ is:

\begin{equation*}
    \begin{array}{cccccc}
Run & A & B & C & D & \\
1   & - & - & - & - & \\
2   & + & - & - & + & \\
3   & - & + & - & + & \\
4   & + & + & - & - & \\
5   & - & - & + & + & \\
6   & + & - & + & - & \\
7   & - & + & + & - & \\
8   & + & + & + & + & \\
    \end{array}
\end{equation*}
\begin{center}
$\vdots$
\end{center}
\end{outputBoxGemini}

\end{document}